\documentclass[%
preprint,
showpacs,
showkeys,
preprintnumbers,
 amsmath,amssymb,
 aps,
 prd,
]{revtex4-1}

\usepackage{graphicx}
\usepackage{dcolumn}
\usepackage{bm}

\begin{document}

\preprint{}

\title{Phenomenology of electroweak bubbles and gravitational waves in the Littlest Higgs Model with T parity}

\author{Sahazada Aziz}
\email{aziz$_$bu@rediff.com}
\author{Buddhadeb Ghosh}
\email{ghoshphysics@yahoo.co.in}
\affiliation{Center for Advanced Study, Department of Physics, University of Burdwan, Burdwan-713104, India.}%

\begin{abstract}
We study the dynamics of electroweak bubbles in the scenario of a strong first order inverse electroweak phase transition at the TeV scale involving the global structure of the nonlinear sigma field in the littlest Higgs model with T parity. Employing the one-loop order finite temperature effective potential, we find that the pressure in the symmetric phase i.e., inside the bubble is always greater than that in the asymmetric phase i.e., outside the bubble, so that the bubbles are expanding. By calculating the fluid velocities in the two phases we arrive at the condition of a supersonic deflagrated motion of the bubble walls. We then discuss the generation of gravitational waves from the collisions of such bubbles as well as from the turbulence of the plasma.
\end{abstract}

\pacs{98.80.Cq, 12.15.Ji, 04.30.-W.}

\keywords{Early Universe; Electroweak phase transition; Beyond Standard Model.}

\maketitle
\section{Introduction}
In recent years, dynamics of electroweak bubbles, associated with a first order phase transition, has been studied in the Standard Model (SM) [1-11] and its extensions [12-15], Two-Higgs Doublet Model (TDHM) [16], the Minimally Supersymmetric Standard Model (MSSM) [17, 18] as well as in model-independent way [19,20]. Also, aspects of gravitational waves (GW), as would be generated by bubble collisions and due to turbulence in the plasma have been investigated [21-28].      
      
Knowledge of bubble dynamics and the concomitant CP violation [29] help us to understand the electroweak phase transition (EWPT) and electroweak baryogenesis (EWBG) scenario in more detail in a specific model.

Depending on the strength of the first-order phase transition and the nature of the finite-temperature effective potential (FTEP) in a model calculation, the motion of the bubble wall may belong to various categories, viz., deflagrations, detonations, hybrids or runaway [20]. Also, the motion can be supersonic, Jouguet or subsonic depending on whether the velocity of the plasma inside the bubble is greater than, equal to or less than the sound velocity in the medium, respectively. Usually, deflagrations are subsonic and detonations are supersonic. However, no clear-cut classification can be made in this regard. Although hydrodynamical equations give general features of bubble wall motion, microscopic analysis including friction of the bubble wall is necessary for getting the detailed information regarding the bubble wall velocity [13, 20].

Gravitational waves, which are considered to be the essential features of Einstein's General Theory of Relativity, have not been experimentally detected so far, although efforts are on in this direction [28]. Gravitational waves at the electroweak scale are believed to have been generated either by bubble wall collisions or by turbulence of the plasma. 

The littlest Higgs model ($L^2$HM) [30] and the littlest Higgs model with T parity (LHT)[31-33] are economical beyond Standard Models (BSM) which can solve the little hierarchy problem. The LHT conforms to the electroweak precision data. In order to explore the prospects of   EWPT and EWBG, finite-temperature calculations have been performed [34-36] in these models. In Ref. 36, an inverse strong first order EWPT has been observed in the global structure of the effective potential at the value of the physical Higgs field, $h=1.1$ TeV with transition temperature, $T_c=0.9$ TeV. 

In Ref. 37, a two-step baryogenesis scenario associated with the inverse phase transition in the TeV scale has been presented.  In the first step, the Universe makes a transition from an electroweak broken phase above $T_c$ to a symmetric phase below $T_c$. Bubbles of symmetric phase are formed in the background of asymmetric phase. Baryon number violations take place within these bubbles due to sphaleron transitions induced by T even massless gauge boson fields. In the second step, a cross-over takes place at $T\approx0.1$ TeV and the process of baryogenesis gets completed between the temperatures, 0.9 TeV and 0.1  TeV.

In view of the proposed new aspects [36] of the EWPT at the TeV scale and the associated EWBG scenario [37], we expect to find new features of the bubble dynamics as well as GW in the LHT.

The purpose of this paper is to examine the properties of the symmetric phase bubbles associated with an inverse strong first order EWPT and a new baryogenesis scenario therewith. In section II, we demonstrate the expansion of the symmetric phase bubbles. In section III, we calculate the bubble nucleation rate and time and in section IV, we study the bubble wall velocity. In section V we investigate the features of gravitational wave generation in our model. Finally, in section VI we make some concluding remarks.

\section{Expansion of the symmetric phase bubbles}
The presence of bubbles in the model under consideration is schematically shown in Fig.1. Here we have a situation which is opposite to that in the case of SM, viz., we have the symmetric phase bubbles in the background of the asymmetric phase. Whether these bubbles will contract or expand will depend on the difference of pressure in the two phases. The pressures are completely determined by the FTEP [36] and can be written as [4],
\begin{equation}
{p_{in}} =  - {V_{eff}}\left( {h = 0,T} \right), \\
{p_{out}} =  - {V_{eff}}\left( {h = 1.1\,TeV,T} \right),
\end{equation}
where, $p_{in}$ and $p_{out}$ are the pressures inside and outside the bubbles respectively. It should be noted that $h=1.1\,TeV$ is the value of the physical Higgs field, where the phase transition is observed.	

\begin{figure}[ht]
\includegraphics[scale=1]{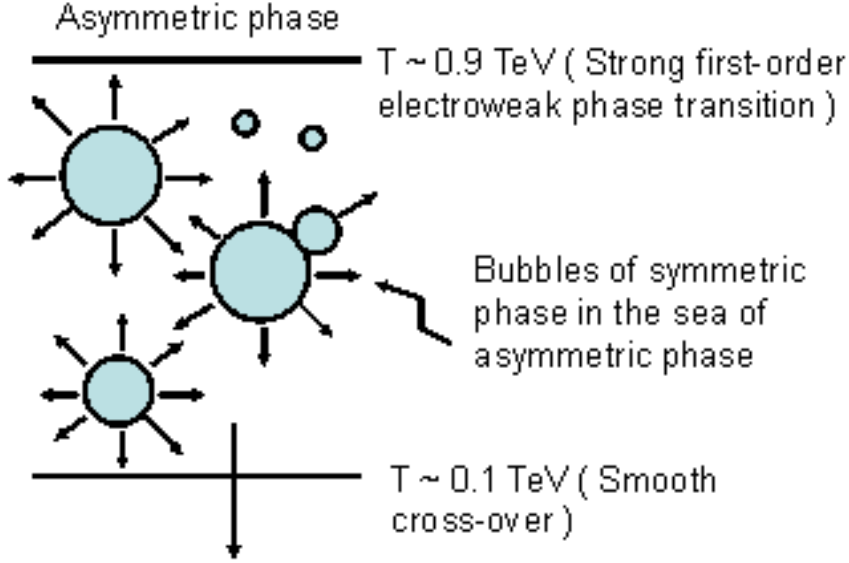}
\caption{Electroweak bubbles in the inverse phase transition scenario in the littlest Higgs model with T parity. Inside the bubbles, the VEV of the physical Higgs field,  $<h>=0$. Outside the bubbles, $<h>=1.1$ TeV.}
\label{fig:Fig.1}
\end{figure}  
So far as the FTEP is concerned, two comments are in order here. Firstly, in the calculation of FTEP in Ref. 36, the effects of mirror fermions, which are instrumental in the non-minimal CP violation [38] in the LHT models, were not taken. Subsequently, we have found that the inclusion of these fermions in the FTEP does not change the features of the phase transition - the value of $T_c$ changes from 0.925 TeV to 0.905 TeV and the position of the minimum of the Higgs field remains the same. However, to be accurate, for the pressure calculations here, we have included the mirror fermions in the FTEP. The value of the mirror fermion coupling constant $\kappa$ has been taken to be 0.7 consistent with a reasonable value of the mirror fermion mass, viz., 0.5 TeV [39]. The second comment concerns the recently discovered [40, 41] SM Higgs of mass, $m_H \simeq 125$ GeV at the LHC. This translates into value of Higgs quartic coupling constant,
 $\lambda  = \frac{{{{\left( {{m_H}/v} \right)}^2}}}{2} \cong 0.13$, $v$ being the SM Higgs VEV, 246 GeV. With this value of $\lambda$, we get a set of UltraViolet (UV) completion factors as $a=-0.01$, $a'= -0.0002$ for our FTEP, consistent with the experimental value of the SM Higgs VEV at zero temperature. Here the UV completion factors are defined as the quantities which take care of the physics above the cut-off scale, $\Lambda  \approx 4\pi f \approx 10\, TeV$, where $f$ is the high energy symmetry breaking scale. It may be noted here that the general procedure for obtaining the UV completion factors in our model is described in Ref. 36.
\begin{figure}[ht]
\includegraphics[scale=1]{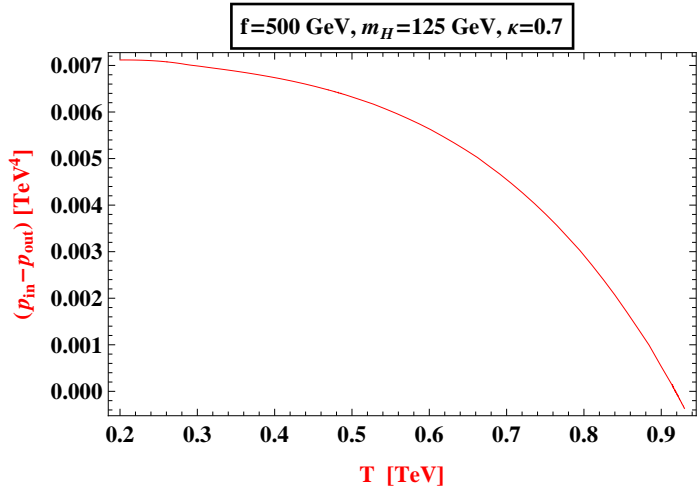}
\caption{Excess pressure within the bubbles of the symmetric phase as a function of temperature.}
\label{fig:Fig.2}
\end{figure}
 
With the above considerations and Eq.1 we have calculated the quantity, ($p_{in}-p_{out}$) of which the plot is shown in Fig.2. We observe that the excess pressure within the symmetric phase bubbles is always positive. From this we may conclude that after the inverse EWPT at $T_c$=0.9 TeV, the symmetric phase bubbles will expand in the background of the asymmetric phase.

\section{Nucleation rate and temperature}
The excess free energy [3] in a true vacuum bubble, which is a symmetric phase bubble in the present case, can be written as,
\begin{equation}
{\rm{\Delta }}F(T) = 4\pi \int_0^R {{r^2}\left[ {\frac{1}{2}{{\left( {\frac{{dh}}{{dr}}} \right)}^2} + V(h,T)} \right]dr},
\end{equation}
where, we have assumed that the bubble is spherical and $R$ is the radius of the bubble. ${{\Delta }}F(T)$ is the same as the three-dimensional Euclidean action, $S_3$ [42] which is related to the four-dimensional Euclidean action $S_E$ as $S_3={S_E}/T$ in the imaginary time formalism of finite-temperature field theory. Eq.2 is the energy which causes the symmetric phase bubble to expand in the present scenario. The derivative term in Eq.2 corresponds to a surface energy and the potential term a volume energy. In the thin wall approximation, appropriate for a strong first-order phase transition, the derivative of the Higgs field can be written as [42], $ \pm \sqrt {2V\left( {h,T} \right)} $ in the limit of exact degeneracy of the potential minima at  $T=T_c$. A positive value of the derivative would reflect the fact that in our case as we go from the centre of a symmetric phase bubble to the boundary, the VEV of $h$  increases from 0 to 1.1 TeV. To avoid using an ansatz of the   field in the $r$ space, we then change the integration variable from $r$ to $h$ and approximate the surface term as, $4\pi {R^2}\sigma (T)$, where $\sigma (T) = \int_{h = 0}^{h = 1.1} {\sqrt {2V\left( {h,T} \right)} {\rm{}}dh}$ can be called the surface energy density. We also write the volume term in Eq.2 as $\frac{4}{3}\pi {R^3}\bar V(h,T)$ where $\bar V(h,T)$ is an average effective potential inside the bubble. Since the potential inside the bubble should be measured with respect to that outside, we can write $\bar V(h,T) = \,{V_{in}} - {V_{out}} =  - {p_{in}} - ( - {p_{out}}) =  - ({p_{in}} - {p_{out}}) =  - \Delta p$. Thus, we write the volume term as, $ - \frac{4}{{3}}\pi {R^3}{\rm{\Delta }}p(T)$.

Now, we get the radius of a critical bubble by minimizing ${\rm{\Delta }}F\left( T \right)$ from the equation, ${\left[ {\frac{{\delta {\rm{\Delta }}F(T)}}{{\delta R}}} \right]_{R = {R_c}}} = 0$ with ${\rm{\Delta }}F(T) = 4\pi {R^2}\sigma (T) - \frac{4}{3}\pi {R^3}{\rm{\Delta }}p(T)$ and thus ${R_c}(T) = \frac{{2\sigma (T)}}{{{\rm{\Delta }}p(T)}}$.

The bubble nucleation rate per unit time per unit volume can be written in terms of the excess free energy $\Delta {F_{C}}(T)$ of the critical bubble as [43,18],
\begin{equation}
{{\rm{\Gamma }}_N}\left( T \right) \simeq {T^4}{\left[ {\frac{{\Delta {F_C}(T)}}{{2\pi T}}} \right]^{3/2}}exp\left[ { - \frac{{\Delta {F_C}(T)}}{T}} \right]
\end{equation}
The bubble nucleation temperature $T_N$,  which is somewhat lower than  $T_c$ in the case of first order phase transition, is defined as the temperature at which the rate of nucleation of a critical bubble within a horizon volume is equal to the Hubble parameter at that temperature. Since the horizon scale is approximately $H{(T)^{ - 1}}$ the nucleation temperature can be defined by the equation,
\begin{equation}
\frac{{{\Gamma _N}({T_N})}}{{{H^3}({T_N})}} \simeq H({T_N}) \simeq 1.66\sqrt {{g_*}({T_N})} \frac{{T_N^2}}{{{m_{Pl}}}}
\end{equation}
where  $g_* (T)$  is the relativistic degrees of freedom and $m_{Pl} \simeq (1.22 \times {10^{16}}TeV)$ is the Planck mass.
From Eqs. (3) and (4), we get,
\begin{equation}
\frac{{\Delta {F_C}({T_N})}}{{{T_N}}} - \frac{3}{2}\ln \left[ {\frac{{\Delta {F_C}({T_N})}}{{{T_N}}}} \right] - 143.46 + 2\ln \left[ {{g_*}({T_N})} \right] + 4\ln {T_N} \simeq 0
\end{equation}
\begin{figure}[ht]
\includegraphics[scale=1]{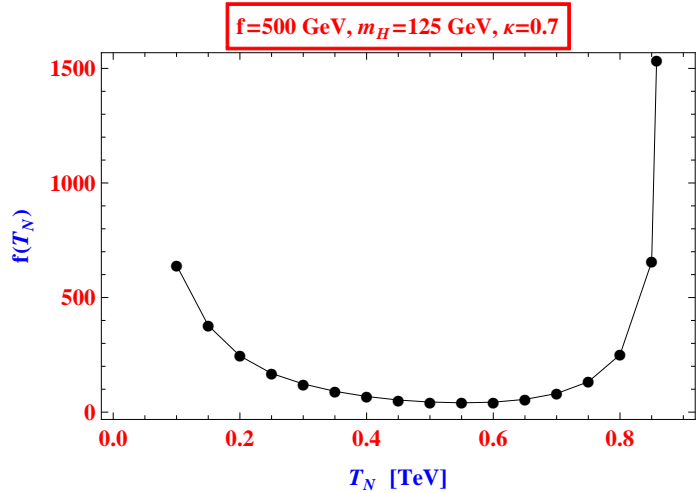}
\caption{A plot of  $f(T_N)$ vs. $T_N$ where $f(T_N)$ is the function on the left hand side of Eq.5 and $T_N$ is the bubble nucleation temperature.}
\label{fig:Fig.3}
\end{figure}  
Eq.5 can be  graphically solved for $T_N$. Denoting the LHS of Eq.5 as  $f(T_N )$,  we  show in Fig.3 a plot of   $f(T_N )$ vs. $T_N$. The graph shows that $f(T_N )\simeq 0$  for  $T_N \sim 0.5-0.6$ TeV.
The nucleation temperature can also be determined from the consideration of the fact that the space-time integrated nucleation rate (i.e., the nucleation probability) should be of order 1[13]:
\begin{equation}
\int_{{t_C}}^{{t_N}} {{\Gamma _N}(t){V_C}dt \sim 1}
\end{equation}
Now the horizon scale is roughly $~2t$ and so, volume = $(2t)^3$. Using the time-temperature relations [42, 13],
\begin{eqnarray}
t = 0.301g_*^{ - 1/2}\frac{{{m_{Pl}}}}{{{T^2}}},\nonumber \\ 
\frac{{dT}}{{dt}} = - HT
\end{eqnarray}
we get the integral in Eq.6, which is the total nucleation probability of a single bubble, as,
\begin{equation} 
I = {\left( {0.602g_*^{ - 1/2}{m_{Pl}}} \right)^4}{\left( {2\pi } \right)^{ - 3/2}}\mathop \smallint \limits_{{T_N}}^{{T_c}} {T^{ - 13/2}}{\left[ {\Delta {F_C}(T)} \right]^{3/2}}exp\left[ { - \Delta {F_C}(T)/T} \right]dT
\end{equation}
Results of evaluation of this integral for different values of $T_N$ are shown in Table 1.
\begin{table}
\caption{Nucleation temperature, excess free-energy and the nucleation probability (integral I in Eq.8).}
	\label{tab:Table 1}
\begin{center}
\begin{tabular}{| c | c | c |}
\hline
$T_{N}$(TeV)  & $\Delta {F_C}(T_N)$ & value of integral (I) \\ \hline \hline
0.55 & 100.91 & $1.37 \times {10^{10}}$\\
0.60 & 111.53 & $1.07 \times {10^5}$\\
0.61 & 114.43 & $4.35 \times {10^3}$\\
0.62 & 117.72 & 113.87\\
0.63 & 120.94 & 3.22\\
0.635 & 122.76 & 0.429\\
0.64 & 124.56 & 0.058\\
0.65 & 128.63 & $6.45 \times {10^{ - 4}}$\\
0.66 & 133.04 & $4.88 \times {10^{ - 6}}$\\
0.67 & 137.98 & $2.05 \times {10^{ - 8}}$\\
0.70 & 156.32 & $3.07 \times {10^{ - 17}}$\\
\hline
\end{tabular}
\end{center}
\end{table}
We see from Table 1 that consistent with Eq.6, the value of $T_N$ should be 0.635 TeV.
 
\section{The nature of bubble and wall velocity}
The entropy, energy and enthalpy densities are defined in terms of the pressure $p(T)$ as,
\begin{eqnarray}
s(T)=(dp(T))/dT \equiv p'(T),\nonumber \\
\rho(T)=Ts(T)-p(T), \nonumber \\
\omega \left( T \right) = \rho \left( T \right) + p\left( T \right) = TS(T).
\end{eqnarray}
We have then the following equations, valid for a first-order phase transition.
\begin{eqnarray}
{p_ + }\left( {{T_c}} \right) = {p_ - }({T_c}), \nonumber \\
{s_ + }\left( {{T_c}} \right) \ne {s_ - }({T_c}), \nonumber \\
{\rho _ + }\left( {{T_c}} \right) \ne {\rho _ - }({T_c}), \nonumber \\
\rm{and} \,\,{\omega _ + }\left( {{T_c}} \right) \ne {\omega _ - }\left( {{T_c}} \right),
\end{eqnarray}
where, the quantities with the subscript, `+' refer to those in the high-temperature phase ( i.e., the phase outside the bubble) and the subscript, `$ - $' is for the quantities in the low-temperature phase (i.e., the phase inside the bubble).
 
The latent heat $L$ is defined as the difference between the energy densities of the two phases at $T=T_c$  and thus,
\begin{equation} 
L = {T_c}[{p'_ + }\left( {{T_c}} \right) - {p'_ - }\left( {{T_c}} \right)]
\end{equation}
Solving hydrodynamical equations [20,44,45], we get,
\begin{eqnarray}
{\omega _ + }\gamma _ + ^2{v_{ + }} = {\omega _ - }\gamma _ - ^2{v_{ - }}, \nonumber \\
{\omega _ + }\gamma _ + ^2v_ + ^2 + {p_ + } = {\omega _ - }\gamma _ - ^2v_ - ^2 + {p_ - }
\end{eqnarray}
where, $\gamma  = 1/\sqrt {1 - {v^2}} $, $v$  being  the velocity being assumed to be aligned in a particular direction, say, the  $z$-direction.

Equivalently, we get,
\begin{eqnarray}
{v_{ + }}{v_ - } = \frac{{{p_ + } - {p_ - }}}{{{\rho _ + } - {\rho _ - }}},\,
\frac{{{v_ + }}}{{{v_ - }}} = \frac{{{\rho _{ - }} + {p_ + }}}{{{\rho _ + } - {p_ - }}}
\end{eqnarray}
whence, we obtain,
\begin{eqnarray}
{v_ + } = \sqrt {\frac{{\left( {{p_ + } - {p_ - }} \right)\left( {{\rho _{ - }} + {p_ + }} \right)}}{{\left( {{\rho _ + } - {\rho _ - }} \right)\left( {{\rho _ + } - {p_ - }} \right)}}}, \nonumber \\        
{v_ - } = \sqrt {\frac{{\left( {{p_ + } - {p_ - }} \right)\left( {{\rho _ + } - {p_ - }} \right)}}{{\left( {{\rho _ + } - {\rho _ - }} \right)\left( {{\rho _{ - }} + {p_ + }} \right)}}}
\end{eqnarray}
where, it is implied that, $p_+=p_+ (T_+ )$,  $p_-=p_- (T_-)$ etc., $T_+$ being the temperature outside the bubble and $T_-$ inside.

The bubble motions can be broadly classified as [13],

(i)	Deflagration : The velocity of the plasma inside the bubble is more than outside, $v_{- }>v_{+ }$ ,

(ii)	Detonation : The velocity outside is more than inside, $v_{+ }>v_{- }$.

Deflagration may further be subdivided as, strong: $v_- >c_s$ , Jouguet: $v_- =c_s$ and weak: $v_- <c_s$ . Here, $c_s$ is the velocity of sound inside the bubble, which has the value $1/\sqrt 3 $ for a relativistic plasma [13]. The strong deflagration is therefore supersonic and the weak deflagration is subsonic. A detonation will be called strong if $v_- <c_s$ , Jouguet if  $v_- =c_s$ and weak if $v_- >c_s$ .
 
Deflagrations and detonations are also characterized by the following temperature constraints [13, 19, 20]:
\begin{eqnarray}
\rm{Deflagration:}\, T_+>T_N>T_- ,\nonumber \\     
\rm{Detonation:}\,   T_->T_+=T_N . 
\end{eqnarray}
The above characterizations show that the detonation cases are easier to investigate because of the condition, $T_+=T_N$.

Now, coming to the case of LHT, let us first assume that the bubble motion is detonated, so that $T_+=T_N$=0.635 TeV. Fig.4 shows the result of calculation of $v_+$ and $v_-$ as a function of $T_-$ for the parameter space $f=500$ GeV, $m_H$ =125 GeV, $\kappa=0.7$  and $T_+=T_N$ (= 0.635 TeV). The plot clearly shows that $v_- >c_s>v_+ $, which is in contradiction with the velocity characterization of detonation. Hence we can rule out the case of detonation in the case of LHT. Also as $v_- >c_s$ in Fig.4, we are getting an indication of  \textit{supersonic deflagration} [19] for the bubble wall motion here.
\begin{figure}[ht]
\includegraphics[scale=1]{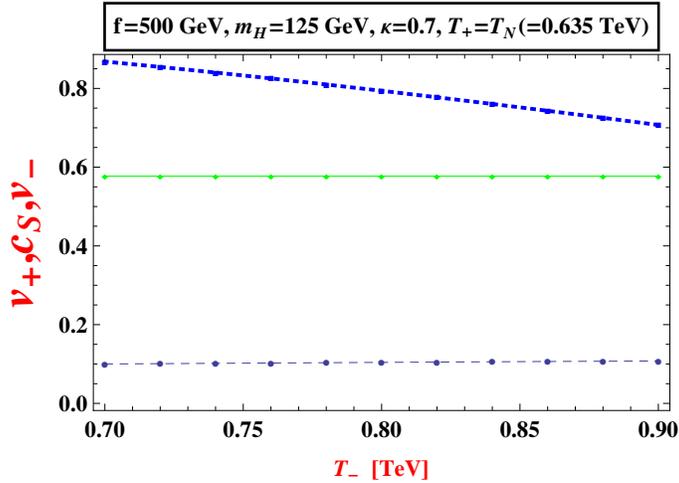}
\caption{Assumming detonation motion (${T_ + } = {T_N}$ ), plot of velocity of plasma outside  the  bubble: ${v_ + }$(lower plot: blue-dashed), sound velocity: ${c_S}( = \frac{1}{{\sqrt 3 }})$, middle green straight line) and velocity of plasma inside the bubble: ${v_ - }$( upper plot: blue-dotted).}
\label{fig:Fig.4}
\end{figure}

To study the bubble wall motion, let us use the following so-called bag equations of state [13]:
\begin{eqnarray}
{\rho _ + } = {a_ + }{T^4} + \varepsilon,\, {p_ + } = \frac{1}{3}{a_ + }{T^4} - \varepsilon,\, {\rho _ - } = {a_ - }{T^4},\,{p_ - } = \frac{1}{3}{a_ - }{T^4}
\end{eqnarray}
and the entropy density,
\begin{equation}
{s_ \pm } = \frac{4}{3}{a_ \pm }{T^3},
\end{equation}
where,
\begin{equation}
{a_ \pm } = \frac{{{\pi ^2}}}{{30}}{g_*}({T_ \pm })
\end{equation}
are numbers related to the number of relativistic species present in the plasma. It is interesting to see the physical significance of the quantity $\varepsilon $. Since at $T_c$, the pressure difference between the two phases is zero, we get from Eq.16,
\begin{equation}
\left( {{a_ + } - {a_ - }} \right)T_c^4 = 3\varepsilon
\end{equation}     
Then, from the expressions of entropy and latent heat we get,
\begin{equation}
L = 4\varepsilon 
\end{equation}
which shows that $\varepsilon $ is closely related to the amount of latent heat $L$ of phase transition.

We now define two important quantities upon which the bubble wall velocity depends:
\begin{eqnarray}
\alpha  = \frac{\varepsilon }{{{a_ + }T_ + ^4}},\, \,{\alpha _N} = \frac{\varepsilon }{{{a_ + }T_N^4}}.     
\end{eqnarray}
From hydrodynamics, we get a relationship between the plasma velocities in terms of $\alpha$ as [46],
\begin{equation}
{{v_ + } = \frac{{\frac{1}{{6{v_ - }}} + \frac{{{v_ - }}}{2} \pm \sqrt {{{\left( {\frac{1}{{6{v_ - }}} + \frac{{{v_ - }}}{2}} \right)}^2} + {\alpha ^2} + \frac{2}{3}\alpha  - \frac{1}{3}} }}{{1 + \alpha }}}
\end{equation}
where, the $+(-)$ sign corresponds to the detonated (deflagrated) motion of the bubble.
As we have argued earlier, the motion of the bubble is deflagrated rather than detonated. Accordingly, we shall study, henceforth, the deflagrated motion of the bubble, taking the `$ - $' sign in Eq.22.
  
A deflagrated bubble proceeds via the generation of shock waves. The expressions of the velocities $v_1$ and $v_2$ of the fluid before and behind respectively of the shock front in the frame of the shock front are [13],
\begin{eqnarray}
\left| {{v_1}} \right| = \frac{1}{{\sqrt 3 }}\left( {\frac{{3T_ + ^4 + T_N^4}}{{3T_N^4 + T_ + ^4}}} \right),\,
\left| {{v_2}} \right| = \frac{1}{{3\left| {{v_1}} \right|}}
\end{eqnarray}
In the laboratory frame the fluid within the bubble, i.e., behind the bubble wall is at rest, as the bubble is isotropically expanding, and also ahead of the shock front, where the temperature is $T_N$, which belongs neither to the low-temperature nor to the high temperature phase, the fluid is at rest. Thus, the wall velocity in the laboratory frame,
\begin{equation}
{v_w} = - {{v_ - }}, {v_{sh}} = - {{v_1}},
\end{equation}
because, $v_-$ is the velocity of the fluid inside the bubble in the reference frame of the wall. In Eq.24 $v_{sh}$ denotes the velocity of the shockfront. Thus in terms of the bubble wall velocity also, it can be said that the motion is supersonic if we can show that $v_- > c_s$ for all temperatures $T_+$ outside the bubble. 
The fluid velocity $v_f$ between the two fronts can be written either in terms of $v_1$, $v_2$ or $v_+$, $v_-$:
\begin{equation}
{v_f} = \frac{{{v_2} - {v_1}}}{{1 - {v_1}{v_{2}}}} = \frac{{{v_ + } - {v_ - }}}{{1 - {v_ + }{v_{ - }}}}
\end{equation}
Putting the values of  $v_1$ and $v_2$ we get,
\begin{equation}
\frac{{{v_ + } - {v_ - }}}{{1 - {v_ + }{v_{ - }}}} = \frac{{\sqrt 3 ({\alpha } - \alpha_ N )}}{{\sqrt {3\left( {{\alpha _N} + \alpha } \right)(3\alpha  + {\alpha _N})} }}
\end{equation}
Eliminating $v_+$ from  Eqs.25 and 26 we get an expression of $v_-$ in terms of $\alpha$ and ${\alpha _N}$:
\begin{eqnarray}
\begin{array}{l}
{v_ - } = {(\sqrt 3 ({\alpha _ + } - {\alpha _N})(3{\alpha _ + } + {\alpha _N})({\alpha _ + } + 3{\alpha _N}))^{ - 1}}\\
\left( \begin{array}{l}
\sqrt {(3{\alpha _ + } + {\alpha _N})({\alpha _ + } + 3{\alpha _N})} (\alpha _ + ^2 - 2{\alpha _ + }(1 + 4{\alpha _ + }){\alpha _N} + \alpha _N^2)\\
 + 2\sqrt (
(3{\alpha _ + } + {\alpha _N})({\alpha _ + } + 3{\alpha _N})(8\alpha _ + ^3\alpha _N^2 + \alpha _N^4 - 2\alpha _ + ^2\alpha _N^2(1 + 2{\alpha _N})\\
 + \alpha _ + ^4(1 + 4{\alpha _N}( - 1 + 4{\alpha _N})))
\end{array} 
 \right)
\end{array}
\end{eqnarray}
In LHT, ${g_*}(T \ge 500\,GeV) \simeq 214$, $T_N$=0.635 TeV, $T_c$=0.91 TeV, $L$=0.2464 TeV. Then we get, ${\alpha _N} = 0.00538.$. 
 
Keeping the latent heat and therefore ${\alpha _N}$ constant, we calculate $v_- $ varying $\alpha$. A plot of $v_-$ against $\alpha$(Fig.5) shows that the motion of the fluid inside the bubble, and therefore the bubble wall velocity $v_w$ in the laboratory frame is supersonic.
\begin{figure}[ht]
\includegraphics[scale=1]{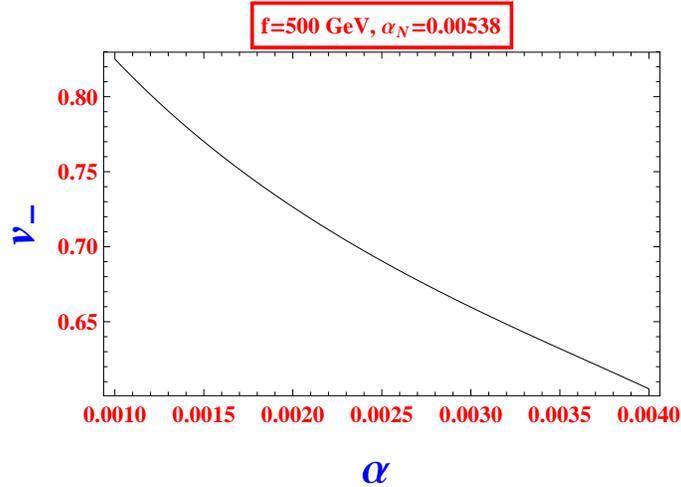}
\caption{Plot of $v_-$ against $\alpha$  for fixed $\alpha _N$ following Eq.27. For various values of the temperature  $T_+$,  $v_-$ is always greater than the sound velocity, $c_s$=$1/\sqrt 3 $=0.577.}
\label{fig:Fig.5}
\end{figure}
\begin{figure}[ht]
\includegraphics[scale=1]{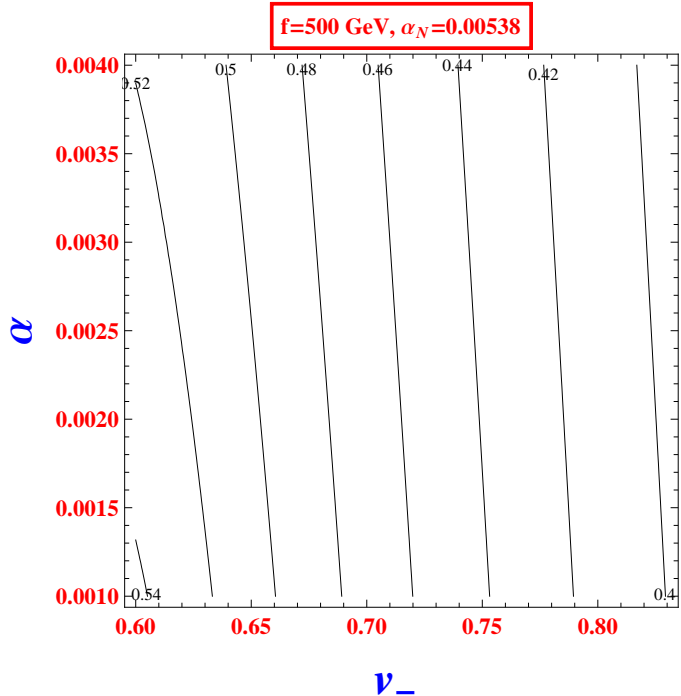}
\caption{Contour plot of $v_-$ against $\alpha$ for various values of $v_+$. Contours denote the values of $v_+$.}
\label{fig:Fig.6}
\end{figure}
We can also plot $v_-$ against $\alpha$ for various values of $v_+$ following Eq.22 instead of Eq.27 where $v_+$ has been eliminated. Such graphs are shown in Fig.6. It is seen from these graphs that for various values of $v_+$ also the supersonic nature of wall velocity i.e., $v_w=|v_- | >c_s$ is maintained.
 
\section{Gravitational waves}
Gravitational waves [48-50] may be generated at the time of the first-order EWPT by two mechanisms [21-28]: (i) Collision of bubble walls, (ii) Turbulence created by the bubble expansion. Since the strong first-order phase transition is not possible in the SM, GW cannot be generated within the framework of the SM.

The frequency and energy density of GW produced in the early Universe change due to the expansion of the Universe. The frequency varies as $a^{-1}$ and the energy density $a^{-4}$ where $a$ is the scale factor. Thus the red-shifted frequency we observe today is given by,
\begin{equation}
{f_0} = {f}\frac{{{a}}}{{{a_0}}},
\end{equation}
where, $a_0$ is the scale factor at the present time and $a ( f )$ is the scale-factor (frequency) at the early Universe, i.e. at the time of the origin of the GW.

Now, because of the adiabatic expansion of the Universe, we can write,
\begin{equation}
a^3{g_*}{T^3} = {a_0}^3{g_0}{T_0}^3
\end{equation}
 Substituting, ${T_0} = 2.73K = 2.35 \times {10^{ - 13}}\,GeV$ and ${g_0} = 2$ we get,
\begin{equation} 
\frac{a}{{a_0}} = 6.4 \times {10^{ - 16}}{\left( {\frac{{100}}{{{g_*}}}} \right)^{1/3}}\left( {\frac{{100\,GeV}}{{{T}}}} \right).
\end{equation}
The wavelength of GW in the early Universe should be a fraction of the Hubble size $H^{-1}$. Hence, it is instructive to write $f_0$ in terms of ${f}/{H}$. Thus, we write,
\begin{equation} 
{f_0} = \left( {\frac{{{a}}}{{{a_0}}}} \right)\left( {\frac{{{f}}}{{{H}}}} \right){H},
\end{equation}
where,  $H$ is given by the Friedmann equation,
\begin{equation} 
H^2 = \frac{{8\pi G}}{3}{\rho},
\end{equation}
$\rho$ being the radiation energy density,
\begin{equation} 
{\rho} = \frac{{{\pi ^2}{g_*}T^4}}{{30}}.
\end{equation}
From the above equations, we get a suggestive expression of frequency of GW in the present Universe,
\begin{equation} 
{f_0} = 1.32 \times {10^{ - 5}}{\left( {\frac{{{g_*}}}{{100}}} \right)^{1/6}}\left( {\frac{{{f}}}{{{H}}}} \right)\left( {\frac{{{T}}}{{100\,GeV}}} \right){\rm{Hz}},
\end{equation}
where, we have used the numerical values [42], ${m_{Pl}} = 1.22 \times {10^{19}}\,GeV$ and $1\,GeV^{- 1} = 6.5822 \times {10^{ - 25}}\,{\sec ^{ - 1}}$.

Let us first consider GW generation from bubble wall collisions. During EWPT, the bubble nucleation rate per unit time and unit volume may be written as [28, 51], 
\begin{equation}
{{\rm{\Gamma }}_{\rm{N}}}\left( t \right) = {{\rm{\Gamma }}_0}\left( {{t_i}} \right){e^{\beta (t - {t_i})}},
\end{equation}
where,  $t_i$ is an initial time and $\beta$ is a parameter setting the time and length scales of the phase transition as ${\beta ^{ - 1}}$ and ${v_w}{\beta ^{ - 1}}$ respectively, $v_w$ being the bubble wall velocity. Assuming $f \sim \beta $, we can write from Eq.34,
\begin{equation}
f_0^{coll} = 1.32 \times {10^{ - 5}}{\left( {\frac{{{g_*}}}{{100}}} \right)^{1/6}}\left( {\frac{{{T}}}{{100\,GeV}}} \right)\left( {\frac{\beta }{H}} \right){\rm{Hz}}.
\end{equation}
A recent simulation [21, 28] gives the dependence of the peak frequency, $f_{peak}^{coll}$ on $v_w$ as,
\begin{equation} 
f_{peak}^{coll} = 1.32 \times {10^{ - 5}}\left( {\frac{{0.62}}{{1.8 - 0.1{v_w} + v_w^2}}} \right){\left( {\frac{{{g_*}}}{{100}}} \right)^{1/6}}\left( {\frac{T}{{100\,GeV}}} \right)\left( {\frac{\beta }{H}} \right){\rm{Hz}},
\end{equation}
which is interestingly similar to Eq.36. We may also observe that the peak frequency of GW due to bubble collisions does not depend very sensitively on the bubble wall velocity.
 
The fractional gravitational energy density (or intensity) in the early Universe is defined as,
\begin{equation} 
{{\rm{\Omega }}_{G{W}}} = \frac{{{\rho _{GW}}}}{{{\rho _{to{t}}}}}
\end{equation}
and its relation with the corresponding value $\Omega _{G{W^0}}^{coll}$ at the present Universe is given by,
\begin{equation} 
\Omega _{G{W^0}}^{coll} = \Omega _{G{W}}^{coll}{\left( {\frac{{{a}}}{{{a_0}}}} \right)^4}{\left( {\frac{{{H}}}{{{H_0}}}} \right)^2}
\end{equation}
Using the value of the Hubble constant in the present Universe [42], ${H_0} = 2.1332h \times {10^{ - 42}}GeV(h \simeq 0.7)$, $H^2 = \left( {8{\pi ^3}/90} \right)G{g_*}T^4 = 1.85 \times {10^{ - 38}}{g_*}T^4\,Ge{V^{ - 2}}$ and Eq.30, we get,
\begin{equation} 
{h^2}\Omega _{G{W^0}}^{coll} = 0.684 \times {10^{ - 5}} \times {\left( {\frac{{100}}{{{g_*}}}} \right)^{1/3}}\Omega _{GW}^{coll}.
\end{equation}
Eq.40 gives a relation between the intensities of the gravitational waves in the early Universe and the present Universe. 

The GW energy density in the early Universe is given by [51],
\begin{equation} 
\Omega _{G{W}}^{coll} = {k^2}{\left( {\frac{\alpha }{{1 + \alpha }}} \right)^2}{\left( {\frac{{{H}}}{\beta }} \right)^2}v_w^3,
\end{equation} 
where, the efficiency factor ($k$) is defined by the equation,
\begin{equation} 
{\rho _k} = k\alpha {\rho _{rad}},
\end{equation} 
$\alpha {\rho _{rad}}$ being the latent heat and $\rho _k$ is the kinetic energy density.

We have here,
\begin{equation} 
{\rho _{tot}} = (1 + \alpha ){\rho _{rad}}
\end{equation} 
From Eq.40 we then get,
\begin{equation} 
{h^2}\Omega _{G{W^0}}^{coll} = 6.84 \times {10^{ - 6}}{\left( {\frac{{100}}{{{g_*}}}} \right)^{1/3}}{k^2}{\left( {\frac{\alpha }{{1 + \alpha }}} \right)^2}{\left( {\frac{{{H}}}{\beta }} \right)^2}v_w^3
\end{equation} 
A recent fit gives [21, 28],
\begin{equation} 
{h^2}\Omega _{G{W^0}}^{coll} \simeq 1.1 \times {10^{ - 6}} \times {\left( {\frac{{100}}{{{g_*}}}} \right)^{1/3}}{k^2}{\left( {\frac{\alpha }{{1 + \alpha }}} \right)^2}{\left( {\frac{{{H}}}{\beta }} \right)^2}\left( {\frac{{v_w^3}}{{0.24 + v_w^2}}} \right)
\end{equation}
which  is close to Eq.44.
 
The GW produced by the stirring of the plasma or by the turbulent bulk motion of the plasma is somewhat different from that due to bubble collisions. In this case the relevant length scale is the so-called `stirring scale' [28], which is approximately given by ${L_s} \approx 2{R_b}$,  where $R_b$ is the bubble radius. For the largest bubble, ${R_{b}} \approx {v_w}{\beta ^{ - 1}}$. A better approximation [28] for this radius is
\begin{equation} 
{R_{b}} \approx 3{v_w}\ln \left( {\frac{\beta }{H}} \right){\beta ^{ - 1}}.
\end{equation} 
As frequency, $f \sim L_s^{ - 1}$, we get from Eq.36,
\begin{equation} 
{f_{peak}^{turb}} \simeq 6.4 \times {10^{ - 6}}{\left( {\frac{{{g_*}}}{{100}}} \right)^{1/6}}\left( {\frac{{{T}}}{{100\,GeV}}} \right)\left( {\frac{1}{{{H}{R_b}}}} \right){\rm{Hz}}
\end{equation}
For the case of GW produced by turbulence, the  expression of the peak intensity of the wave can be cast in the form [28, 52-55],
\begin{equation} 
{h^2}\Omega _{GW}^{turb} \simeq 3.5 \times {10^{ - 5}}{\left( {\frac{{100}}{{{g_*}}}} \right)^{1/3}}{\left( {{k} \frac{\alpha }{{\left( {1 + \alpha } \right)}}} \right)^{3/2}}\left( {\frac{{{{({R_b}H)}}}}{{1 + 4\frac{{3.5}}{{{{({R_b}H)}}}}}}} \right)
\end{equation}
We now turn to the calculations of the frequency and intensity of GW in our model. We have written the bubble nucleation rate both as functions of temperature (Eq.3) and time (Eq.35). The rates are equal at time and temperature related by Eqs.7. From Eq.35, we get,
\begin{equation} 
\beta  = \frac{{{{{\rm{\dot \Gamma }}}_N}}}{{{{\rm{\Gamma }}_N}}}
\end{equation} 
Then, from Eqs.7 and 3 we obtain,
\begin{equation}
\frac{\beta }{H} = T\frac{d}{{dT}}\left( {{\rm{\Delta }}{F_{C}}(T)/T} \right)
\end{equation} 
where, we have neglected the time variation of the prefactor in comparison to that of the exponent as the nucleation rate is mainly driven by the exponential factor [28].

To get the derivative $T\frac{d}{{dT}}\left( {{\rm{\Delta }}{F_C}(T)/T} \right)$ at various temperatures in the desired range we have made a polynomial fit of  ${{\rm{\Delta }}{F_C}(T)/T}$ up to sixteenth power of  $T$ from its values calculated at a number of temperatures and tabulated in Table 2.
\begin{table}
\caption{Values of various quantities, which are useful for studying bubble dynamics.}
	\label{tab:Table 2}
\begin{center}
\begin{tabular}{| c | c | c | c | c |}
\hline
$T(TeV)$ & $\sigma (Te{V^3})$ & ${R_{C}}(Te{V^{ - 1}})$ & $\Delta {F_C}(TeV)$ & $\Delta {F_C}/T$ \\ \hline \hline
0.10 & 0.06198 &  17.48 & 79.28 & 792.87\\
0.15 & 0.06197 &  17.46 & 79.15 & 527.68\\
0.20 & 0.06196 &  17.41 & 78.70 & 393.50\\
0.25 & 0.06187 &  17.44 & 78.90 & 315.62\\
0.30 & 0.06170 &  17.66 & 80.63 & 268.77\\
0.35 & 0.06153 &  17.89 & 82.54 & 235.85\\
0.40 & 0.06130 &  18.19 & 85.02 & 212.56\\
0.45 & 0.06102 &  18.60 & 88.48 & 196.63\\
0.50 & 0.06067 &  19.19 & 93.59 & 187.19\\
0.55 & 0.06023 &  20.00 & 100.91 & 183.48\\
0.60 & 0.05964 &  21.12 & 111.53 & 185.88\\
0.65 & 0.05883 &  22.84 & 128.63 & 197.89\\
0.70 & 0.05779 &  25.41 & 156.32 & 223.31\\
0.75 & 0.05634 &  29.53 & 205.89 & 274.52\\
0.80 & 0.05435 &  37.14 & 314.21 & 392.76\\
0.85 & 0.05145 &  56.13 & 679.04 & 798.87\\
0.90 & 0.04667 &  172.06 & 5786.28 & 6429.20\\
\hline
\end{tabular}
\end{center}
\end{table}
The fitted polynomial is,
\begin{eqnarray}
\Delta {F_C}(T)/T =  63631.9 - 2.88719 \times {10^6}T + 6.01409 \times {10^7}{T^2}- 7.54798 \nonumber \times {10^8}{T^3}\\ 
 + 6.39515 \times {10^9}{T^4} - 3.88944 \times {10^{10}}{T^5} + 1.7616 \times {10^{11}}{T^6} - 6.07722 \times \nonumber {10^{11}}{T^7}\\ 
 + 1.61766 \times {10^{12}}{T^8} - 3.34018 \times {10^{12}}{T^9} + 5.34127 \times {10^{12}}{T^{10}} - 6.55427 \times {10^{12}}{T^{11}}\\ \nonumber
 + 6.05712 \times {10^{12}}{T^{12}} - 4.07896 \times {10^{12}}{T^{13}} + 1.88892 \times {10^{12}}{T^{14}} - 5.37703 \times {10^{11}}{T^{15}}\\ \nonumber
 +7.09076 \times {10^{10}}{T^{16}}.
\end{eqnarray}
Differentiating the function (Eq.51) with respect to $T$ we can get the value of $\beta /H$ at a particular temperature.
 
In Ref.20 an expression  of efficiency factor for supersonic  deflagration has been derived in model-independent way:
\begin{equation} 
{k = {k_1} + ({v_W} - {c_S}) \times \delta  + {{\left( {\frac{{{v_W} - {c_S}}}{{{c_J} - {c_S}}}} \right)}^3}({k_2} - {k_1} - ({c_J} - {c_S}) \times \delta )},
\end{equation}
\begin{equation} 
{k_1} = \frac{{{\alpha ^{2/5}}}}{{0.017 + {{(0.997 + \alpha )}^{2/5}}}},\,{k_2} = \frac{{\sqrt \alpha  }}{{0.135 + \sqrt {0.98 + \alpha } }},\,\delta  =  - 0.9\ln \left( {\frac{{\sqrt \alpha  }}{{1 + \sqrt \alpha  }}} \right)
\end{equation}
and  $c_S$  and $c_J$ are the sound and Jouguet velocities respectively.
\begin{equation} 
{{c_J} = \frac{{\sqrt {1/3}  + \sqrt {{\alpha ^2} + 2\alpha /3} }}{{1 + \alpha }}}
\end{equation}
In Fig.7, we show the variation of GW signal generated by bubble wall collisions against temperature as well as the bubble wall velocity. The plot shows that the intensity is practically independent of the bubble wall velocity. This behavior can be understood from the $v_w$ dependence of the intensity, as demonstrated by Eq.45. This feature of velocity independence has been found here in the case of frequency also as we have already mentioned. 
\begin{figure}[ht]
\includegraphics[scale=1]{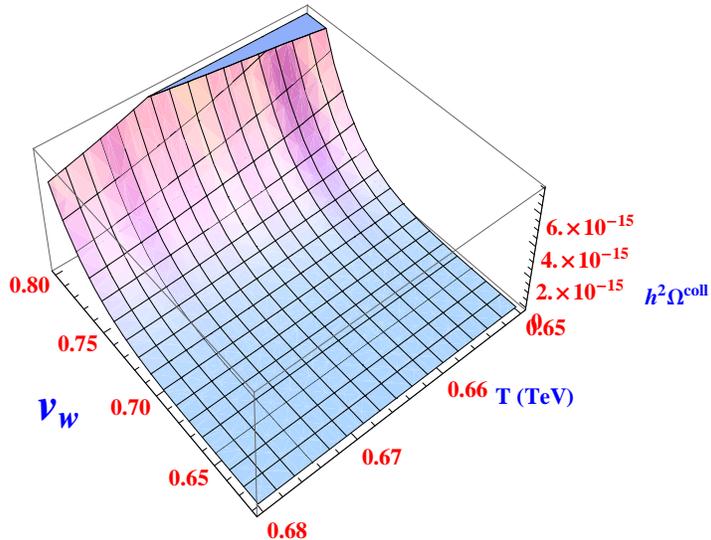}
\caption{Plot of variation of intensity of GW from bubble wall collisions against wall velocity $v_w$  and  temperature $T$.}
\label{fig:Fig.7}
\end{figure}
Next, we have compared the GW intensities for the cases of bubble collisions and turbulence for a fixed value of the wall velocity ($v_w$=0.7) and for temperatures above the nucleation temperature $T_N$=0.635 TeV, upto temperature near  $T_C$=0.91 TeV in Fig.8. In the same figure we have plotted the value of ${(\beta /H)^{ - 2}}$. The trend of the GW signal from bubble collisions and turbulence follows the variation of the quantity ${(\beta /H)^{ - 2}}$. As the temperature is lowered from $T_C$, more and more bubbles are formed which eventually increases the likelihood of more bubble collisions. Thus the intensity of GW increases as the temperature approches $T_N$. Similarly, as the radii of the bubbles increase with decrease in temperature [see Eq.46] the stirring scale $L_s$ of turbulence increases making the motion of the plasma more turbulent and thus the intensity of GW due to turbulence is enhanced. We also note that the signal from turbulence to be more enhanced than that from bubble collisions. Specifically, the value of the  peak intensity arising out of turbulence is about $10^{-13}$ whereas that due to bubble collisions is $10^{-16}$ at temperure around $T_N$. The calculation shows that the turbulent motion of the plasma was more instrumental in producing GW in the early Universe than the bubble collisions.
\begin{figure}[ht]
\includegraphics[scale=1]{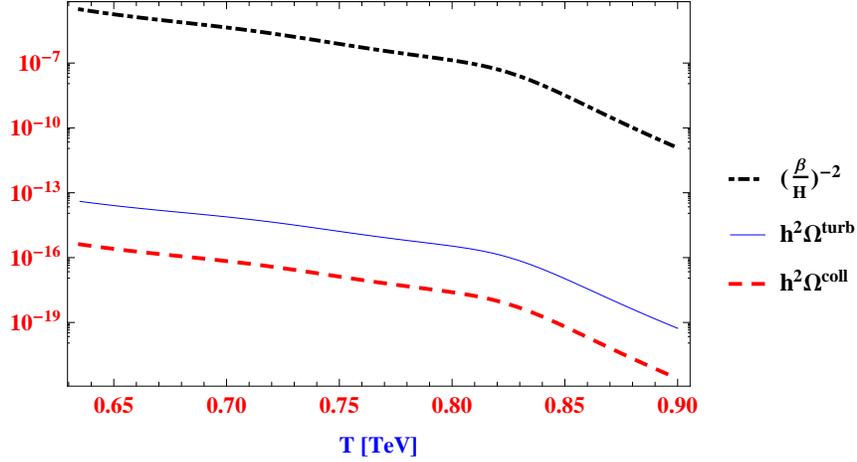}
\caption{Plot of Intensity of GW due to turbulence (solid line) and from wall collisions (dashed line). Bubble wall velocity is held fixed at an intermediate value, $v_W = 0.7$. Upper plot shows the variation of ${(\beta /H)^{ - 2}}$ with temperature $T$.}
\label{fig:Fig.8}
\end{figure}
\begin{figure}[ht]
\includegraphics[scale=1]{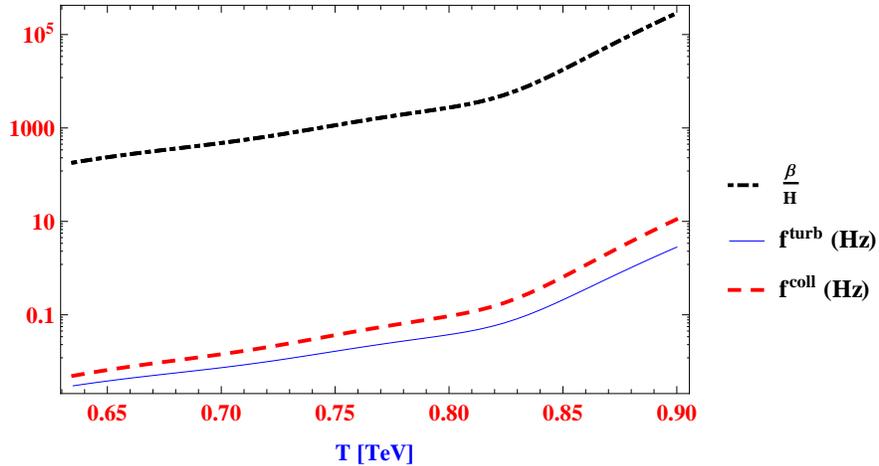}
\caption{Plot of frequency of GW due to bubble collisions (dashed line) and from turbulence (solid line). Bubble wall velocity is held fixed at an intermediate value, $v_W = 0.7$. Upper plot (dash-dotted line) shows the variation of $\frac{\beta }{H}$ with temperature $T$.}
\label{fig:Fig.9}
\end{figure}

In Fig.9 we have shown the frequency distribution against temperature for GW from turbulence as well as bubble collisions. We have also plotted the variation of $\frac{\beta }{H}$ with temperature. The figure shows that the peak freuencies of GW are essentially proportional to the values of $\frac{\beta }{H}$. The peak frequencies around the region of peak intensities are only fractions of a Hertz.
The very low value of GW intensity obtained in our analysis can be understood by noting that the intensity is roughly given by (see Eqs. 45 and 48),
\begin{equation} 
{\Omega _{GW}} \approx {10^{ - 5}}{k ^2}{\alpha ^2}{\left( {\frac{\beta }{H}} \right)^{ - 2}}
\end{equation}

In our model, we have got small value of $\alpha \,( \sim {10^{ - 3}})$ and large value of $\beta /H\,( \sim {10^3})$ (obtained from Eqs.50 and 51). The low value of $\alpha$ is because of two reasons (see Eq. 21): (i) large number of relativistic species (=214, more than double of the SM value) present in the plasma and (ii) high temperature of the plasma ${\rm{(T}} \sim {\rm{TeV)}}$.
   
On the other hand, the GW frequency is roughly given by (see Eqs. 37 and 47)
\begin{equation}
{f_{GW}} \approx {10^{ - 5}}\left( {\frac{\beta }{H}} \right)\,{\rm{Hz}}
\end{equation}   
which explains the value of peak frequency in the deciHz range.
    
The weak GW signals that are obtained in our calculations are difficult to detect in the ongoing GW detectors including LIGO. However, there are good possibilities of detection in Ultimate DECIGO [28] and BBO Correlated [48], as the sensitivities of these detectors conform to the frequency and intensity range obtained in the present calculation. These distinct signals of the gravitational waves are characteristic signatures of the high temperture non-standard electroweak phase transition explored [36] in LHT.  
\section{Conclusions}
   In conclusion, we have studied in the present paper some aspects of the dynamics of the bubbles associated with first-order electroweak phase transition within the framework of the littlest Higgs model with T parity. We had earlier, in this model noticed a new region of strong first-order phase transition and proposed a baryogenesis scenario at the TeV scale. Our general observation in the present work is that the bubble wall motions are supersonic deflagrated. It may be noted that the supersonic expansion of symmetric phase bubbles has been considered recently [14] in a different baryogenesis model.  We have also considered the generation of gravitational waves due to bubble collisions as well as turbulence in the plasma. As expected, the calculated intensities for the latter case are three order of magnitude larger than for the former case, at their peak values. Although the obtained intensities are quite small in magnitude, we expect them to be detected by some of the future gravitational wave detectors. 

\section*{Acknowledgments}
One of the authors (S.A.) acknowledges the Council of Scientific and Industrial Research, the Government of India for granting a Senior Research Fellowship, under which part of the work was done. He also thanks Ariel M$\acute{e}$gevand for illuminating correspondences and Jos$\acute{e}$ M. No for valuable comments.  B.G. thanks  Patrick Dasgupta of Delhi University for useful discussions.

\end{document}